\documentclass{aa}
\usepackage[varg]{txfonts}
\usepackage{multirow}
\usepackage{amsmath}
\usepackage{bm}
\usepackage{enumerate}
\usepackage{booktabs}
\usepackage{orcidlink}
\usepackage{lscape}
\usepackage{graphicx,wrapfig,lipsum}

\title{Investigating the influence of radio-faint AGN activity on the infrared-radio correlation of massive galaxies}

   \author{Giorgia Peluso
          \fnmsep\thanks{Email: giorgia.peluso@inaf.it}
          \orcidlink{0000-0001-5766-7154}
          \inst{1}
          \and
          Ivan Delvecchio 
          \fnmsep\thanks{Email: ivan.delvecchio@inaf.it}
          \orcidlink{0000-0001-8706-2252}
          \inst{1} 
         \and
         Jack Radcliffe 
         \orcidlink{0000-0002-0813-0497}
          \inst{2,3}
         \and
         Emanuele Daddi
         \orcidlink{0000-0002-3331-9590}
          \inst{4}
         \and
         Roger Deane 
         \orcidlink{0000-0003-1027-5043}
          \inst{3,5}
         \and \\
         Matt Jarvis 
         \orcidlink{0000-0001-7039-9078}
          \inst{6}
         \and
         Giovanni Zamorani
         \orcidlink{0000-0002-2318-301X}
          \inst{1}
         Isabella Prandoni 
         \orcidlink{0000-0001-9680-7092}
          \inst{7}
         \and
         Myriam Gitti
         \orcidlink{0000-0002-0843-3009}
          \inst{7,8}
         \and
         Cristiana Spingola
         \orcidlink{0000-0002-2231-6861}
          \inst{7}
         \and
         Francesco Ubertosi
         \orcidlink{0000-0001-5338-4472}
          \inst{7,8}
         \and
         Mark Sargent
        \orcidlink{0000-0003-1033-9684}
          \inst{9}
         \and
         Vernesa Smol\v{c}i\'c
         \orcidlink{0000-0002-3893-8614}
          \inst{10}      
         \and
          Wuji Wang
         \orcidlink{0000-0002-7964-6749}
          \inst{11}
         \and
         Jacinta Delhaize
        \orcidlink{0000-0002-6149-0846}
          \inst{12} 
         \and
         Shuowen Jin
        \orcidlink{0000-0002-8412-7951}
          \inst{13,14}
         \and
         Adam Deller
         \orcidlink{0000-0001-9434-3837}
          \inst{15}
        }

   \institute{
   INAF $-$ Osservatorio di Astrofisica e Scienza dello Spazio di Bologna, via Gobetti 93/3, I$-$40129, Bologna, Italy
   \and 
   Jodrell Bank Centre for Astrophysics, University of Manchester, Oxford Road, Manchester M13 9PL, UK
   \and
   Department of Physics, University of Pretoria, Lynnwood Rd, Hatfield, Pretoria, 0002, South Africa
   \and
   CEA, Universit\`e Paris-Saclay, Université Paris Cit\`e, CNRS, AIM, 91191, Gif$-$sur$-$Yvette, France
   \and
   University of the Witwatersrand, Enoch Sontonga Ave, Johannesburg, South Africa
   \and
   University of Birmingham, Birmingham B15 2TT, United Kingdom
   \and
   INAF $-$ Istituto di Radioastronomia, via Gobetti 101, I$-$40129, Bologna, Italy
   \and 
   Dipartimento di Fisica e Astronomia (DIFA), Universit\`a di Bologna, via Gobetti 93/2, I-40129 Bologna, Italy
   \and
   Institute of Physics, Laboratory of Astrophysics, École Polytechnique Fédérale de Lausanne (EPFL), 1290 Sauverny, Switzerland
   \and
   Department of Physics, University of Zagreb, Bijeni\v{c}ka cesta 32, Zagreb, Croatia
   \and
   Caltech/IPAC, 1200 E. California Blvd. Pasadena, CA 91125, USA
   \and
   University of Western Australia, 35 Stirling Hwy, Crawley WA 6009, Australia
   \and
   Cosmic Dawn Center (DAWN), Denmark
   \and
   DTU Space, Technical University of Denmark, Elektrovej 327, 2800 Kgs. Lyngby, Denmark
   \and
   Swinburne University of Technology, John Street, Hawthorn, Victoria, Australia, 3122
      }

\date{2025}
\titlerunning{A\&A,}
\authorrunning {Peluso, G., et al.}

\begin{document}

\abstract
{It is well-known that star-forming galaxies (SFGs) exhibit a tight correlation between their radio and infrared emissions, commonly referred to as the infrared-radio correlation (IRRC). 
Recent empirical studies have reported a dependence of the IRRC on the galaxy stellar mass, in which more massive galaxies tend to show lower infrared-to-radio ratios ($q_{IR}$) with respect to less massive galaxies.
One possible, yet unexplored, explanation is a residual contamination of the radio emission from active galactic nuclei (AGN), not captured through "radio-excess" diagnostics.
}
{To investigate this hypothesis, we aim to statistically quantify the contribution of AGN emission to the radio luminosities of SFGs located within the scatter of the IRRC.}
{Our VLBA program "AGN-sCAN" has targeted 500 galaxies that follow the q$_{IR}$ distribution of the IRRC, i.e., with no prior evidence for radio-excess AGN emission based on low-resolution ($\sim$ arcsec) VLA radio imaging. Our VLBA 1.4 GHz observations reach a 5$\sigma$ sensitivity limit of 25 $\mu$Jy/beam, corresponding to a radio brightness temperature of $T_b \sim 10^5$ K. This classification serves as a robust AGN diagnostic, regardless of the host galaxy's star formation rate.}
{We detect four VLBA sources in the deepest regions, which are also the faintest VLBI-detected AGN in SFGs to date. The effective AGN detection rate is 9\%, when considering a control sample matched in mass and sensitivity, which is in good agreement with the extrapolation of previous radio AGN number counts. Despite the non-negligible AGN flux contamination ($\sim$30\%) in our individual VLBA detections, we find that the peak of the $q_{IR}$ distribution is completely unaffected by this correction. Although we cannot rule out a high incidence of radio-silent AGN at (sub)$\mu$Jy levels among the VLBA non-detections, we derive a conservative upper limit of $<0.1$ dex to their cumulative impact on the $q_{IR}$ distribution. We conclude that residual AGN contamination from non-radio-excess AGN is unlikely to be the primary driver of the M$_\star$ - dependent IRRC.
}{}

\maketitle

\section{Introduction}

Star-forming galaxies show a positive relation between total infrared (L$_{IR}$, rest-frame 8-1000 $\mu$m) and monochromatic radio luminosity (e.g. at 1.4 GHz; L$_{1.4}$), which keeps remarkably tight \citep[$\sigma \sim$ 0.26 dex; e.g.][]{bell2003} and linear for over three orders of magnitude in both $L_{\rm{IR}}$ and L$_{\rm{1.4 GHz}}$ \citep[e.g.,][even though a mild non-linearity was also found in low-$z$ galaxies, e.g. at $z<0.2$ \citealt{molnar+2021}]{helou+1985, condon1992}.

This is generally known as the "infrared-radio correlation" (IRRC).
The origin of the IRRC arises from the relationship between radio luminosity and (obscured) star formation rate (SFR). Broadly speaking, the IR emission arises from the light of massive OB stars, which is absorbed and re-emitted by dust \citep{madau-dickinson2014}, and therefore directly traces the galaxy's SFR. On the other side, radio-continuum emission at GHz frequencies originates mainly from synchrotron processes, produced by relativistic cosmic ray electrons (CRe) accelerated by shock waves produced when massive stars ($> 8 M_\odot$) explode as supernovae, in the case of star-forming galaxies (SFG).

The logarithmic ratio between the two luminosities, labelled "$q_{IR}$", is defined as:

\begin{equation}
    q_{IR} = {\rm \log \left(\frac{L_{IR} [W]}{3.75 \times 10^{12} [Hz]} \right) - \log ( L_{1.4 GHz} [W/Hz])}
    \label{eq:qIR}
\end{equation}

where 3.75 $\times$ 10$^{12}$ Hz represents the central frequency over the far-infrared (e.g., 42-122 $\mu$m) regime.

However, the nature of the cosmic evolution of the IRRC has long been debated \citep[e.g.,][]{harwit_pacini1975, rickard_harvey1984, dejong+1985, helou+1985, hummel+1988, condon1992, garrett2002, appleton+2004, jarvis+2010, sargent+2010, ivison+2010a,ivison+2010b, bourne+2011, magnelli+2015, delhaize+2017, gurkan+2018, molnar+2018, algera+2020b, basu+2015,calistro_rivera+2017}.

Several studies \citep{magnelli+2015, calistro_rivera+2017, delhaize+2017} observed a dependence of the IRRC on the redshift, in the form of $q_{IR} \propto (1 + z)^{\delta}$ with $ -0.1 < \delta < -0.2$. Other works \citep[][D21 hereafter]{gurkan+2018, smith+2021, delvecchio+2021} have recently argued that the IRRC evolves primarily with $M_\star$, with more massive galaxies showing a systematically lower $q_{IR}$, while the dependence on the redshift is secondary and weaker, as indeed $q_{IR} \propto (1 + z)^{-0.023 \pm 0.008}$ at fixed stellar mass \citep[see also,][]{gurkan+2018, smith+2021}.
Among the possible physical drivers of the $q_{IR}$ decline observed in massive star-forming galaxies, the energy loss of cosmic ray electrons (CRe) has been suggested to play a significant role \citep[e.g.][]{schober+2023, yoon+2024}. Specifically, \cite{yoon+2024} modelled the CRe energy-loss as a function of the gas density and redshift of the host galaxy. Their findings indicate an anti-correlation between these quantities: specifically, galaxies with higher gas densities \citep[or, equivalently, higher stellar masses, e.g.][]{kennicutt+1998}, experience less intense energy losses, arguably because they retain more easily their CRes. On the reverse, less massive galaxies are generally most affected by CRe's energy losses. As a consequence of these synchrotron energy losses, less massive galaxies have lower radio luminous with respect to more massive ones.

One yet unexplored explanation for the $M_\star$-evolution is that a hidden and widespread contribution from radio AGN activity might be driving the decline of the $q_{IR}$, especially at high stellar masses ($\log (M_\star/M_\odot) \geq 10$).

One very powerful tool to assess the nature of radio emission is the brightness temperature of the emitter ($T_b$), defined as:

\begin{equation}
     T_b[K] =  1.22 \times 10^6 \times (1+z) \times \left(\frac{S_{obs}}{\mu\rm{Jy}}\right)\left(\frac{\nu}{GHz}\right)^{-2} \left(\frac{\theta_x \cdot \theta_y}{mas^2}\right)^{-1}
     \label{eq:t_b}
\end{equation}

where $S_{obs}$ is the observed density flux at the frequency $\nu$ and $\theta^{x}$ and $\theta^{y}$ are the major and minor axes of the beam \citep{ulvestad+2005}.

Traditionally, high $T_b$ measurements ($T_b \geq 10^5$K) require Very Long Baseline Interferometry (VLBI), which reach milliarcsec resolutions at GHz frequencies \citep[e.g.,][]{middelberg+2011,middelberg+2013,rampadarath+2015, herrera-ruiz+2017, herrera-ruiz+2018, radcliffe+2018, radcliffe+2021, njeri+2023, deane+2024}. Recent studies performed with the Very Long Baseline Array (VLBA) at few milli-arcsec (mas) resolution (e.g., 10 mas $\sim$ 80 pc at z=2) have identified AGN even in galaxies whose emission is dominated by star formation at Very Large Array (VLA) scales \citep{herrera-ruiz+2017, herrera-ruiz+2018, maini+2017}.  
However, these pioneering studies have been highly biased towards radio-excess AGN (i.e., showing excess radio emission relative to that expected from pure star formation; \citealt{smolcic+2017}.
Our project 'AGN-sCAN' extends this type of analysis by targeting with the VLBA a representative sample of SFGs without strong VLA radio-excess, down to a sensitivity reaching as deep as $\sigma\approx$5 $\mu$Jy/beam at 1.4 GHz, at high angular resolution (average beam size: 26$\times$7 mas$^2$). From Eq.~\ref{eq:t_b}, the $T_b$ limit at $>$5$\sigma$ for an unresolved source at z=0.5 is $T_b \geq 10^5$~K \footnote{This value is calculated for the elongated VLBA beam in the COSMOS field (0.027$^{\prime\prime}$x0.007$^{\prime\prime}$, at Dec=2${\deg}$), hence this limit differs, at fixed depth, from that of other VLBA surveys at northern declinations \citep[e.g.GOODS-N][]{deane+2024}.}. Compact radio cores ($\sim$ mas scales) characterized by such $T_b$ would unambiguously arise from an AGN \citep[][]{condon1992,radcliffe+2018, morabito+2022}.

Section \ref{sec:sample} describes the main physical properties of the radio targets composing the sample used in this work; Section \ref{sec:data} summarizes the AGN-sCAN observational strategy and data products, of which a complete and detailed description will be presented by Delvecchio et al. in prep. (D25 hereafter); Section \ref{sec:results} summarizes our main result: the contribution of AGN emission to the nuclear emission of SFGs, classified through the IRRC, is negligible.

We adopt a \cite{chabrier+2003} initial mass function in the mass range 0.1-100 M$_\odot$. We assume a standard $\Lambda$CDM cosmology with $\Omega_m$ =0.3, $\Omega_{\Lambda}$ = 0.7 and $H_0$ = 70 km $\rm s^{-1}$ Mpc$^{-1}$. 
We define the radio spectral index $\alpha$ with the following convention: S$_{\nu} \propto \nu^{- \alpha}$, where $S_{\nu}$ is the flux density at a frequency $\nu$.

\section{Sample selection and physical properties} \label{sec:sample}

This VLBA program (ID: VLBA/20B-140, PI: Delvecchio) has targeted 500 massive (M$_{\star} \geq 10^{10}$ M$_{\odot}$) star-forming galaxies in the Cosmic Evolution Survey (COSMOS, \citep{scoville+2007}), which is an equatorial 2~deg$^2$ field observed with all major space- and ground-based telescopes\footnote{An exhaustive overview of the COSMOS field is available at: \url{http://cosmos.astro.caltech.edu/} and all the relevant datasets for this study are detailed in D25.}. We refer to D25 for the detailed description of the survey.

The parent galaxy sample was retrieved from the COSMOS-Web survey \citep{casey+2023}, containing deep imaging in four JWST/NIRCam filters (F115W, F150W, F277W, F444W) over 0.54~deg$^2$ \citep{franco+2025}, and with JWST/MIRI 7.7$\mu$m over 0.2 deg$^2$ \citep{harish+2025}. Given that our VLBA pointing lies within the COSMOS-Web NIRCam footprint, we update all redshifts and counterpart properties from the latest COSMOS-Web photometric catalogue \citep{shuntov+2025}, which extends in wavelength out to \textit{Spitzer}-IRAC bands. Photometric redshifts are in very good agreement with spectroscopic measurements (i.e. median absolute deviation $\sigma_{MAD}$$\sim$0.012 at m$_{F444W}$$<$28; see \citealt{khostovan+2025}). We adopt the spectroscopic measurement if available and reliable ($>$95\% confidence level) based on the compilation by \cite{khostovan+2025}.

In order to retrieve a q$_{\rm IR}$ measurement for each target, we require our galaxies to have been detected ($\geq$5$\sigma$) at both infrared (IR) and radio frequencies. The combined IR dataset comes from de-blended far-infrared/sub-mm data from \textit{Herschel} (100--500~$\mu$m), SCUBA2 (450, 850~$\mu$m), AzTEC-1.1~mm, MAMBO-1.2~mm \citep{jin+2018}, and ALMA from the recent A$^3$COSMOS survey (\citealt{liu+2019}; \citealt{adscheid+2024}), whenever available. For each source, the total IR luminosity (8-1000$\mu$m, L$_{\rm IR}$) is obtained via spectral energy distribution (SED) fitting, and corrected for potential IR-AGN emission \citep{jin+2018}. Radio counterparts are jointly taken from MIGHTEE 1.28~GHz \citep[6$^{\prime \prime}$ resolution,][]{jarvis+2016,heywood+2022}, VLA 1.4 GHz \citep[1.5$^{\prime \prime}$ resolution,][]{schinnerer+2010} and VLA 3GHz \citep[0.75$^{\prime \prime}$ resolution, ][]{smolcic+2017} imaging. The rest-frame 1.4 GHz luminosity (L$_{1.4}$) is derived from the radio dataset with the highest available S/N, and $K$-corrected to rest-frame 1.4~GHz by assuming a power law spectrum with spectral index $\alpha$ between 3~GHz and 1.28/1.4~GHz, or assuming ${\alpha}$=0.75 if not jointly detected.
To match the spatial resolutions between the 3~GHz and 1.4~GHz images, we downgraded the 3~GHz map by convolving it with a Gaussian kernel of 1.5$^{\prime\prime}$ FWHM \citep[similarly to, e.g.,][]{gim+2019}. We refrain from using a larger FWHM (i.e. for 1.28 GHz counterparts from MeerKAT), as \cite{delhaize+2017} showed that the flux increase above 1.5$^{\prime\prime}$ is typically negligible. The new 3~GHz fluxes at 1.5$^{\prime\prime}$ resolution are higher by 6\%, on average, compared to those of the native resolution (0.75$^{\prime\prime}$) map. This effect induced a minimal flattening in the re-calculated spectral index for the joint 1.4-3 GHz detections \citep[see Sec 4.4.1 in][for a in-depth discussion about the influence of the radio spectral index on the IRRC]{delhaize+2017}.

As for the IR part, we require that the combined S/N (summed in quadrature over all {\it Herschel}, SCUBA2, AzTEC, and ALMA bands) is higher than 5. For the same reason, for the radio part, we require the 1.4~GHz flux to be higher than 25$\mu$Jy, that is 5$\times$ the VLBA 1.4~GHz sensitivity at the pointing center.
We restrict ourselves to galaxies at $z$$>$0.5, in order to get the full galaxy size within the VLBA field of view limited by bandwidth smearing (i.e. 5.5$^{\prime \prime}$ diameter).
We disregard all previously detected VLBA targets from \citeauthor{herrera-ruiz+2017} (\citeyear{herrera-ruiz+2017}; \citeyear{herrera-ruiz+2018}) to avoid duplications.

These selection criteria naturally identify 500 massive (M$_{\star}$$\gtrsim$10$^{10}$~M$_{\odot}$) galaxies.
This sample is representative of star-forming galaxies on the IRRC, as they are broadly distributed in $q_{IR}$ around the $(M_\star,z)$-dependent IRRC from \cite{delvecchio+2021} (see also Sec \ref{sec:results} for details).

\section{Observation and data calibration} \label{sec:data}

Our VLBA observations are summarized in the following, while further details will be given in D25.
In each VLBA session, the full L-band bandwidth is split in 8 spectral windows of 32~MHz each, further divided in 16 spectral channels (2 MHz each) at 4sec time resolution. This setup yields a field of view limited by time (bandwidth) smearing of  10.82$^{\prime\prime}$ (2.75$^{\prime\prime}$), hence larger than the typical galaxy size at z$>$0.5 \citep{yang+2025}.
The final images of all targets, consisting of 4096$\times$4096 pixels (at 0.001$^{\prime\prime}$/px scale) weight about 67 Mbyte each \footnote{The images are publicly available at: \url{https://github.com/idelvecchio/AGN-sCAN_catalogue}}.

The pointing center was located at RA (J2000) = 10:00:36.3299 and Dec (J2000) = +02:12:58.773 and covers 30$^{\prime}$ FWHM at 1.4 GHz. It was observed for a total of 120 hr, split in 24 epochs of 5-hr each, due to the limited visibility of COSMOS by all 10 VLBA antennas for longer tracks. Raw data were calibrated using the VLBI PIPEline \citep[VPIPE;][Radcliffe et al., in prep., \footnote{\url{https://github.com/jradcliffe5/VLBI_pipeline}}]{radcliffe+2024} that is designed to perform automated instrumental calibration, phase referencing, and imaging for multi-phase centre VLBI observations. All these steps have been successfully tested in former very long baseline surveys (EVN from \citealt{radcliffe+2018}; \citealt{radcliffe+2021}; VLBA from \citealt{deane+2024}).

\begin{figure*}[ht]
    \centering
    \makebox[\textwidth]{
    \includegraphics[scale=0.2]{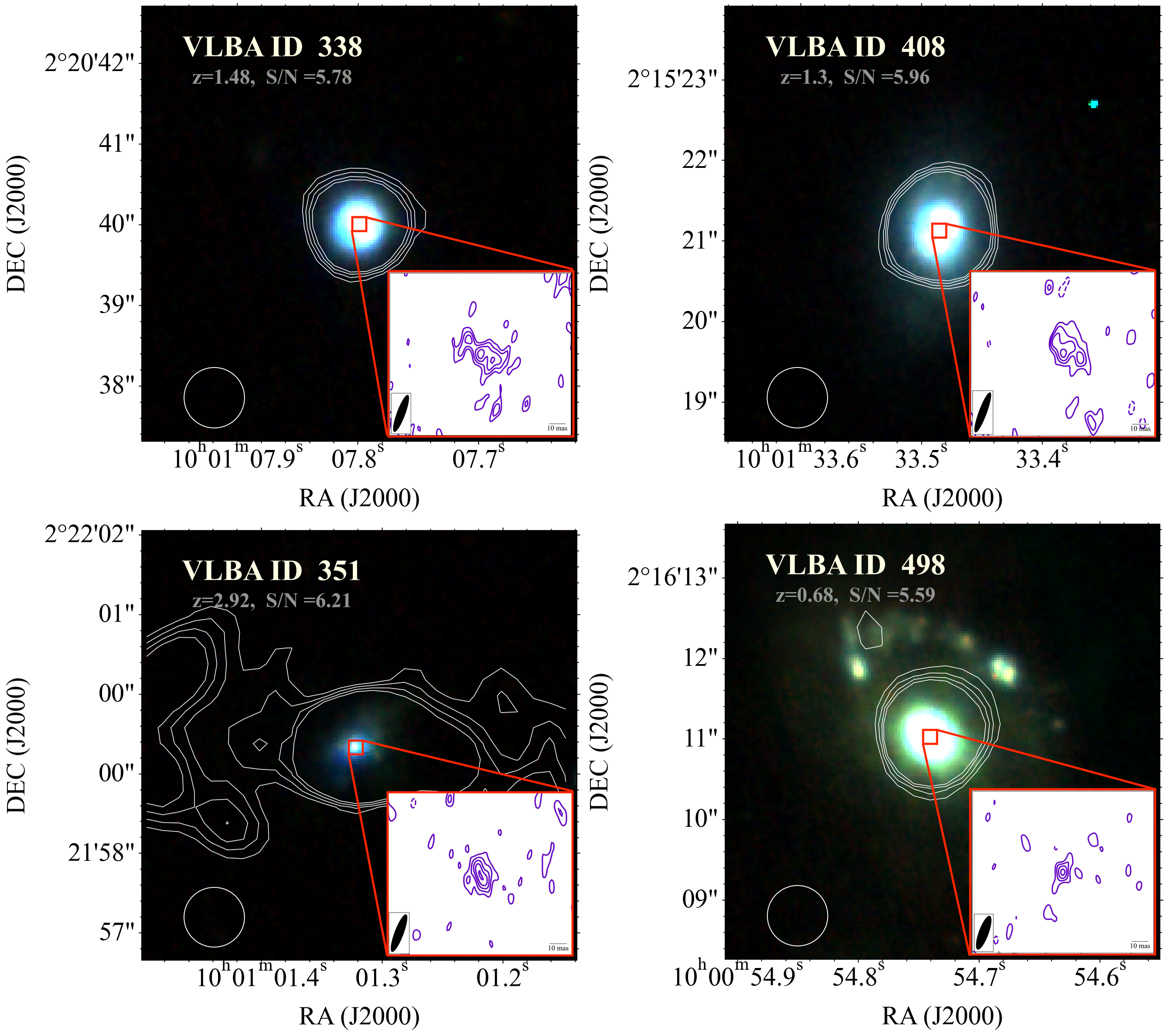}}%
    %VLBA_COSMOS_mask05_image.jpg
        \caption{RGB images of the targets VLBA ID '338' (top left), '351' (top right), '408' (bottom left), and '498' (bottom right). These are obtained by combining the F150W, F277W, and F444W JWST/NIRCam filters \citep{franco+2025} covering a 5$^{{\prime}{\prime}}$$\times$5$^{{\prime}{\prime}}$ field of view. The NIRCam pixel scale is 30~mas. The white contours are the VLA 3~GHz levels \citep{smolcic+2017} at 3, 4, 5, and 6 times the rms (2.3 $\mu$Jy/beam, the VLA beamsize is 0.75$^{{\prime}{\prime}}$ as shown by the white circle in the bottom left). In the zoom-in red panels, the purple contours are the levels at 2, 3, 4, 5, and 6 times the image rms detected by the VLBA, which come from a \(0.2^{\prime \prime} \times 0.2^{\prime \prime}\) field of view, also outlined by the red square at the centers of the galaxies. The dotted purple contour also shows the -3 rms level. The VLBA beam size for each detection is: 26$\times$5 mas$^2$ ('338'), 26$\times$5 mas$^2$ ('351'), 28$\times$5 mas$^2$ ('408'), and 23$\times$6 mas$^2$ ('498'), with a similar position angle PA$\sim$162$^{\circ}$. }
    \label{fig:detections}
\end{figure*}

\begin{table*}[h]
\caption{Summary of the 4 VLBA-AGN's properties}
\footnotesize
\begin{tabular}{cccccccccccccccc} 
 ID$_{\rm VLBA}$ & RA & DEC & $z$ &  $\log (M_\star/M_\odot)$ &  rms$_{\rm VLBA}$ & $S_{\rm VLBA}$ & $\log L_{\rm VLBA}$ & $\log L_{\rm VLA}$ & $\log(L_{\rm IR}$/L$_\odot$) & $q_{IR}$ \\
   & [deg] & [deg] & &  & [$\mu$Jy/beam]  & [$\mu$Jy] & [W/Hz] & [W/Hz] & &  \\
 \\
 \hline
 \hline
 \\
 338  & 150.28249 & 2.34445 & 1.48 &  10.85$\pm$0.04 &  5.6 & 50.2 $\pm 12.9$ & 23.77$\pm$0.13 & 24.19$\pm$0.02 & 12.14$\pm$0.04 & 1.95$\pm$0.04 \\
 351  & 150.25549 & 2.36649 & 2.92 & 10.80$\pm$0.04 &  6.1  &41.4 $\pm 9.3$ & 23.88$\pm$0.11 & 24.80$\pm$0.02 & 12.62$\pm$0.15 & 1.84$\pm$0.15 \\
 408  & 150.38951 & 2.25589 & 1.30  & 11.10$\pm$0.04 &  5.6  &58.2 $\pm 13.4$ & 23.69$\pm$0.11 & 24.20$\pm$0.08 & 11.87$\pm$0.05 & 1.68$\pm$0.09 \\
 498  & 150.22804 & 2.26971 & 0.68 & 10.65$\pm$0.04 & 6.7  &37.2 $\pm  7.8$ & 22.79$\pm$0.10 & 23.32$\pm$0.03 & 11.69$\pm$0.03 & 2.37$\pm$0.04 \\
 \\
 \hline
\end{tabular}
\\
\tablefoot{Columns are: 1) VLBA identification number; 2) and 3) VLBA centroid coordinates; 4) spectroscopic redshifts: $z=1.48$ \citep{kashino+2019} , $z=2.92$ \citep{horowitz+2022}, $z =1.30$ \citep{kartaltepe+2015}, $z=0.68$ \citep{vanderwel+2021}; 5) stellar mass in Chabrier IMF \citep{shuntov+2025}; 6) average VLBA rms ($\mu$Jy/beam); 7) rest-frame VLBA 1.4GHz flux density in $\mu$Jy (S$_{VLBA}$); 8) rest-frame VLBA 1.4GHz luminosity; 9) VLA 1.4GHz rest-frame radio luminosity; 10) total IR luminosity (rest-frame 8-1000 $\mu m$) obtained from SED-fitting with deblended IR fluxes; 11) IR-to-radio luminosity ratio (see Equation \ref{eq:qIR}).}
\label{tab:detections}
\end{table*}

In order to evaluate the significance of each VLBA source as a function of the peak $S/N$, we have run simulations using the Common Astronomy Software Applications (CASA, v.6.6) package. We refer the reader to D25 for more technical details on this task. Briefly, we considered the real noise map, taken from an empty sky region at the VLBA pointing centre, and inject one point-like source in the $uv$-plane, with known position and S/N. We iterated the simulation 100 times, keeping the input position fixed but varying the input S/N over the range 4.5 $\leq$ S/N $\leq$ 7.5. Only one source was injected into each realization, in order to mitigate sidelobe effects and obtain reliable output fluxes (see D25). We then imaged each simulated dataset adopting the same setup used for real data and employing BLOBCAT \cite{hales+2012} to extract the flux and position of all blobs above a given peak S/N, in each image. While the completeness at S/N$>$4.5 is always observed to be higher than 90\% (i.e. BLOBCAT finds the one input source in $>$90\% of runs), the fraction of spurious blobs is a strong function of S/N. In particular, we defined the "spuriousness" ($p_{spur}$) as:

\begin{equation}
    p_{spur} = 1 - \frac{N_{inp}}{N_{out}}
\end{equation}

where $N_{inp}$ is the number of 'real' (input) blobs and $N_{out}$ is the number of the (output) detected blobs. We count the output blobs within a radius of $\pm$0.2$^{\prime \prime}$ from the input (real) source, similarly to what we do for real detections (see Sect.\ref{detections}). 
We find that the spuriousness is higher than 50\% at S/N $<$5, while it decreases to $\approx$15\% at S/N$\sim$5.5, and exponentially drops below 2\% at S/N$\geq$6. If counting all random blobs over the full image (4$^{\prime \prime}\times$4$^{\prime \prime}$), the spuriousness at fixed S/N increases proportionally to the searching area, except at S/N$>$6 where it drops to below 2\% as found in the central area.

\section{Final sample of VLBA detections} \label{detections}

Following \cite{herrera-ruiz+2017} (HR2017, hereafter), we obtain the total fluxes from BLOBCAT \citep{hales+2012} by integrating the blobs with S/N$\geq$5.5 that are found within $\pm$0.2$^{\prime \prime}$, that is $\pm$1 VLA 3~GHz pixel from the prior VLA counterpart position \citep{smolcic+2017}, which sets our maximum VLBA-VLA separation - in HR2017 it was $\pm$0.4$^{\prime \prime}$, i.e. $\pm$1 VLA 1.4~GHz pixel \citep{schinnerer+2010}. For consistency, our chosen size of $\pm$ 0.2$^{\prime \prime}$ is also the circular region inside which we forced the cleaning in order to maximize the S/N (D25).
Based on the above criteria, we identify $4$ VLBA detections associated with AGN activity, with their main properties listed in Table \ref{tab:detections}. We also show them in Figure \ref{fig:detections}, where the presence of a compact radio core is evident (e.g., purple contours). The average VLBA rms of these four detections is 6$\mu$Jy/beam.
Interestingly, these sources are the faintest AGN ever detected with VLBI techniques in SFGs. If pushing our detections threshold at lower S/N, in particular at $5.0 < S/N < 5.5$, we would identify two additional source candidates, but their high spuriousness ($\sim$40\% at their average S/N$\sim$5.2) does not allow us to claim them as secure detections.

We note in Fig.\ref{fig:detections} that the VLBA (iso-surface brightness) contours appear more complex than the elongated synthesized beam. This is because 3 out of 4 detections (all but '498') are marginally resolved (along the beam minor axis), according to BLOBCAT \citep{hales+2012}. However, given the low S/N of these detections, the limited $uv$-coverage and the lack of a-priori information on the source morphology, the imaging and deconvolution process becomes highly uncertain (e.g. \citealt{taylor+1999}). In such conditions, the final restored contours can be a complex blend of different spatial components and are not expected to strictly follow the shape of the clean beam, hence some contours of faint sources would appear rounder and/or smaller than the synthesized beam. This is a known issue in deep VLBI imaging and has been observed in other VLBI surveys (e.g., \citealt{radcliffe+2018}). Because the 3 resolved detections display non-Gaussian morphologies, as recommended in \cite{hales+2012} we do take total VLBA fluxes (e,g, $S_{VLBA}$ in Table \ref{tab:detections}) from the corresponding observed integrated values, but skipping the Gaussian volume correction. For the only unresolved detection ('498'), instead, we set the total flux equal to the peak flux.

In order to make a fair comparison between our VLBA detections and non-detections, we define two control samples of VLBA non-detections ("CS" hereafter): (i) a M$_\star$-matched CS (i.e., with log ($M_\star/M_\odot) >$  10.5; 277 objects) and a combined [$M_\star, rms$]-matched CS (i.e. a subset of the first CS, with also rms $< 6.7 \mu$Jybeam$^{-1}$, i.e. the highest among the 4 detections; 42 objects). Therefore, this latter CS contains the VLBA targets which were not detected within a depth-matched area, despite being evenly accessible as for the VLBA detections.
Overall, the effective detection rate in our sample is 9\% (i.e. 4/46), which is broadly consistent with the AGN duty cycle in massive star-forming galaxies \citep{delvecchio+2022, wang+2022}. We note that matching the control sample by the lowest or the average rms of the detections would not have changed the effective detection rate.

\section{The AGN corrected infrared-to-1.4 GHz radio luminosity ratio} \label{sec:results}

\begin{figure}[ht]
    \centering
    \includegraphics[scale=0.45]{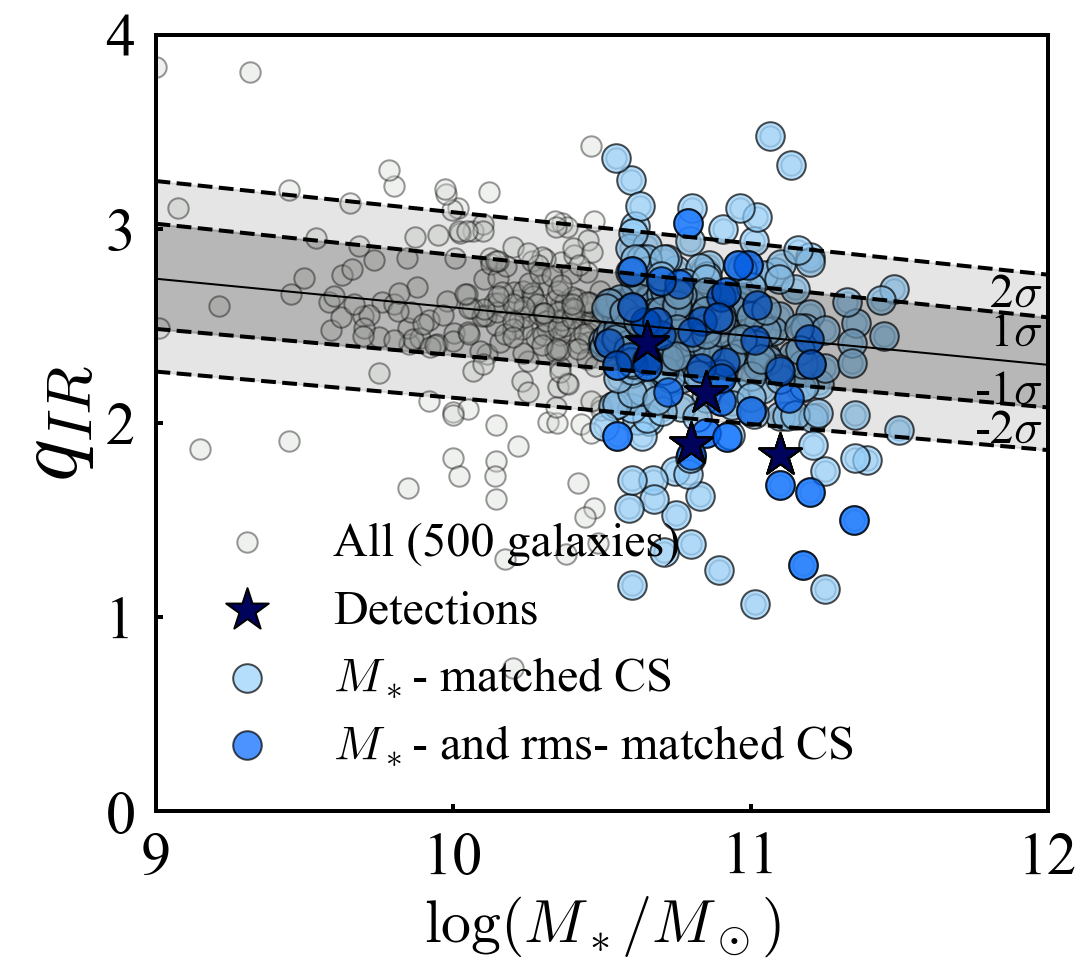}
    \caption{$q_{IR}$ distribution as a function of host galaxy stellar mass for all the 500 targets (grey points), the 4 VLBA-detected AGN (blue stars) and the 277 non-detections part of the $M_\star$-matched sample (light blue points). The $q_{IR}$ of the detections account for the AGN emission detected by the VLBA (e.g., $q_{{\rm IR,corr}}$). The darker blue points are the 42 non-detections observed at sensitivities equal to or lower than the sensitivity of the detections, thus constituting the (M*, rms)- matched CS. 
    The black line is the $q_{IR}$ versus $\log M_\star / M_\odot$ relation from \cite{delvecchio+2021} at $<z> \ \sim 1$, and the shaded areas spans the scatter of the IRRC
    at $M_\star > 10^{10.5}M_\odot$ from \cite{delvecchio+2021}, of around $\pm$ 0.22 dex and $\pm$ 0.44 dex.
    }
   \label{fig:qIR_mass}
\end{figure}

We measured $q_{IR}$ exploiting the total radio VLA luminosities, where both star formation and AGN activity can, in principle, contribute. Figure \ref{fig:qIR_mass} shows these values as a function of the stellar mass of the host galaxy. Dark stars mark the $4$ detections, the light blue circles indicate the $ M_\star$-matched CS, while the darker blue circles represent the ($M_\star, rms$) - matched CS. 
The black solid line represents the dependence of the IRRC on the stellar mass for galaxies at a $ \langle z \rangle \sim 1$, as indicated by the relationship described by \cite{delvecchio+2021}:

\begin{equation}
q_{IR} (M_\star, z) = 2.646 \pm 0.024 \times A^{-0.023 \pm 0.008} - B \times (0.148 \pm 0.013)
\label{eq:qIRvsmass}
\end{equation}

where $A=(1+z)$ and $B = \log (M_\star/M_\odot)$. 
\\

\begin{figure*}[ht]
    \centering
    \includegraphics[scale=0.55]{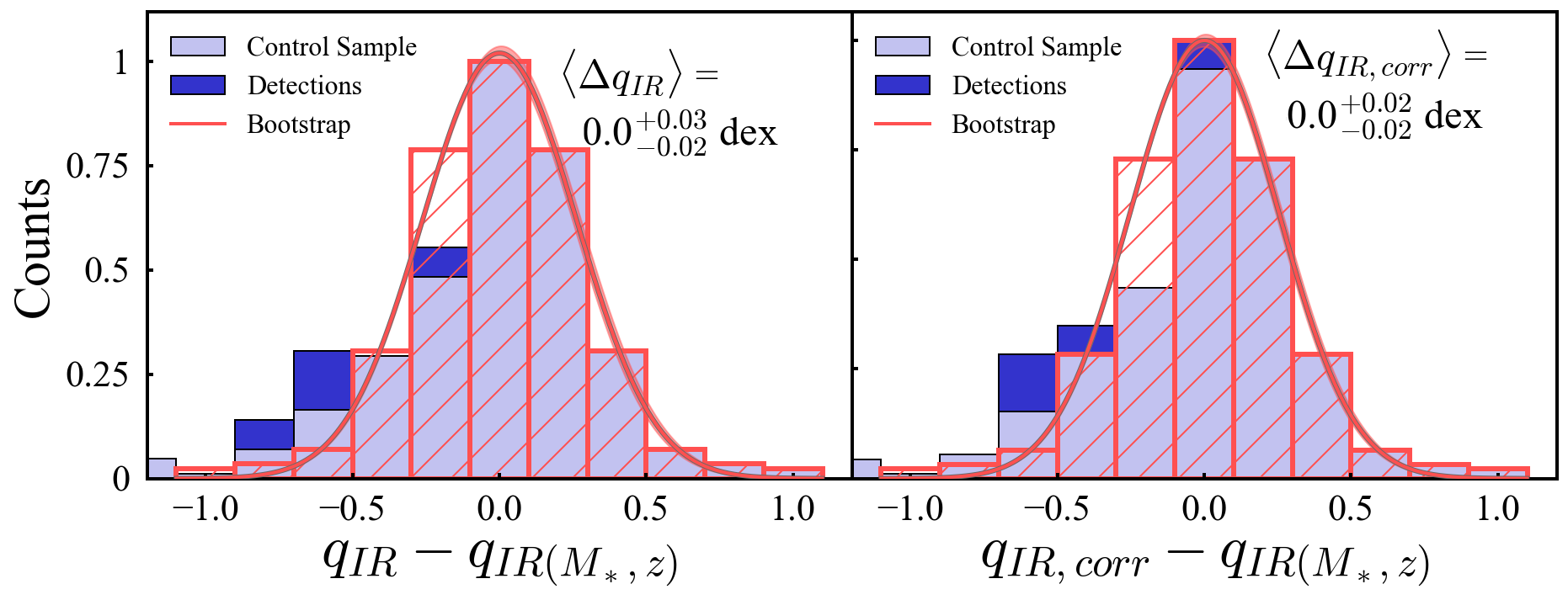}
    \caption{Normalized distributions of the difference between the observed $q_{IR}$ and the one expected from Eq. \ref{eq:qIRvsmass} \citep{delvecchio+2021} based on the stellar mass and the redshift of the host galaxy, $q_{IR}$ (M$_\star$,z).
    The light purple histogram represents the distribution of the (M$_\star$,z)-CS. The stacked dark purple histogram (i.e. dark purple histogram on top of the light purple one) represents the detections, and the hatched orange histogram is the so-called mirror distribution, interpreted as the intrinsic $q_{IR}$ distribution of SFGs. The bootstrapped median values of the mirror distribution were fitted with a Gaussian (red curve; see Sec. \ref{sec:results} for further details).
    Left panel: $q_{IR}$ is computed using the VLA luminosity coming from both star formation and AGN activity. Right panel: $q_{IR,corr}$ is the ratio measured after subtracting the AGN radio luminosity detected by the VLBA from the total radio VLA emission.}
   \label{fig:deltaqir_mass}
\end{figure*}

It is important to note that the dependence on the stellar mass is stronger than that with the redshift.  To measure the $q_{IR} (M_\star, z)$, we consider the stellar masses and redshifts presented in Sec \ref{sec:data}. 
Since the $q_{IR} (M_\star, z)$ distributions are similar between the $M_\star$- and ($M_\star, rms$) - matched CS (as expected given that the observed $q_{IR}$ does not depend on the VLBA rms), in the following we consider the $q_{IR} (M_\star, z)$ distribution of the $M_\star$-matched CS to increase the statistics.

If we assume that the incidence of AGN is as high as in the deepest regions across the entire sky area mapped by the AGN-sCAN survey, we can then reason that the number of effective detections would have been 7 times greater than the $N_{AGN} = 4$ detected in regions with rms $<$6.7 $\mu$Jy/beam. This factor of seven was determined by noting that the subset of non-detections matched on both stellar mass and rms, ($M_\star, rms$)-CS, is approximately seven times smaller than the subset matched solely by stellar mass (e.g., $M_\star$-CS).

Then, we measured the difference ($\Delta q_{IR}$) between $q_{IR} (M_\star,z)$ and the observed $q_{IR}$ computed using the VLA luminosities in Eq. \ref{eq:qIR}. On the left panel of Fig. \ref{fig:deltaqir_mass}, the light purple histogram displays the distribution of $\Delta q_{IR}$ for the (M$_\star$, rms)-CS, while the dark purple histogram displays the distribution of $\Delta q_{IR}$ for the detections. 
To reproduce the expected IRRC for pure SFG, we build up the so-called 'mirror distribution' (orange dashed histogram). This is obtained by mirroring the right side of the real distributions, assumed to be purely originating from SF processes, onto the left side. The errors associated with the counts of the mirror distributions (orange shaded area)  are calculated with a bootstrap, by extracting 1000 values from a Gaussian distribution centered on the observed number counts ($N$), in each bin, with a standard deviation of $ \sqrt{N}$.
We observe that the mirror distribution peaks around 0 in both panels. This allows us to visually assess how much the observed distribution deviates from a perfect Gaussian.  In the right-hand panel of Fig. \ref{fig:deltaqir_mass}, we instead plot the distributions of the difference between $q_{IR} (M_\star,z)$ and the corrected q$_{IR}$ ($q_{IR,{\rm corr}}$), after subtracting the AGN emission observed with the VLBA. In this case, $\Delta q_{IR,\rm{corr}}$ equals $\log [ (L_{VLA}) / (L_{VLA} - L_{VLBA}) ]$,  which is $>0$ by definition. 

The first thing to note is that the median values of the real distributions remain unchanged before and after the correction for the AGN contribution to the radio luminosities. Both before and after the correction, we estimate \( \langle \Delta_{qIR} \rangle = 0.0^{+0.03}_{-0.02} \) dex. The errors on the median are calculated as the 16th and 84th percentiles of the distribution, divided by the square root of the total counts. 
However,  it was expected that the AGN contribution would have minimal impact on the scatter within the IRRC for these galaxies, due to the low fraction of detections.

Nevertheless, it is interesting to note that on an individual source basis, the AGN fluxes substantially influence $q_{IR}$, since the average difference between $q_{IR}$ and $q_{IR,corr}$ for the detections is about 0.14 dex (e.g. 30\%), which is not negligible.

To further explore the potential role of radio variability, we examined the list of variable 1.5~GHz sources identified in the CHILES VERDES survey \citep{sarbadhicary+2021}. The survey provides 1-2 GHz continuum VLA data obtained between 2013 and 2019 in a 0.44 $deg^2$ area of the COSMOS field for 750 previously detected at VLA-1.4GHz. Moderate  variability (10-30\% flux density variation) was observed in 18 sources, and low-variability (2-10\% flux density variation) in 40 sources, for a total of 58/750 (6\%) variable targets. Only one of those variable sources is in common with our VLBA 1.4~GHz targets (even though it is not a VLBA detection) and exhibits low VLA variability ($\pm$5\%).
Moreover, based on the comparison between VLBA-1.4GHz \citep{herrera-ruiz+2017} and VLA-1.4GHz \citep{schinnerer+2010} fluxes taken at different epochs (2012 vs 2007, respectively), \citet{herrera-ruiz+2017} found 13/468 ($\sim$3\%) VLBA detections with VLBA flux higher than the total VLA flux (at above $>$$\pm$1$\sigma$ uncertainty), which the authors interpreted as possible sign of variability. However, those 13 sources all have S$_{\rm vlba}$$>$200~$\mu$Jy, making them among the brightest of their sample. Hence, we expect a negligible fraction of radio-variable sources among our even fainter targets.

Then, we also find that the real distribution exhibits a deviation from the Gaussianity (e.g., when compared to the mirror distribution) on the left side, in both panels. However, this is minimal, as indeed a K-S test could not reject the hypothesis that these two distributions come from the same parent sample (with a p-value of 0.997). \\
Finally, we note that the mirror distributions are basically unchanged even after correcting the $q_{IR}$ of the detections. Besides the peak, also their standard deviations are fully consistent (from $\sigma=\pm 0.44$ dex to $\sigma=\pm 0.45$ dex). Overall, these results highlight that VLBA AGN contamination, despite being non-negligible for individual detections, does not impact at all the peak position (and the overall shape) of the original $q_{IR}$ distribution.

For completeness, in Appendix \ref{sec:appendix-2}, we correct the values of $q_{IR}$ for the dependence on redshift, and we verified that also in that case the distribution of the $z$-corrected $q_{IR}$ distribution remains unchanged before and after the AGN correction.

\section{Source counts}

\begin{figure*}[ht]
    \centering
    \includegraphics[scale=1]{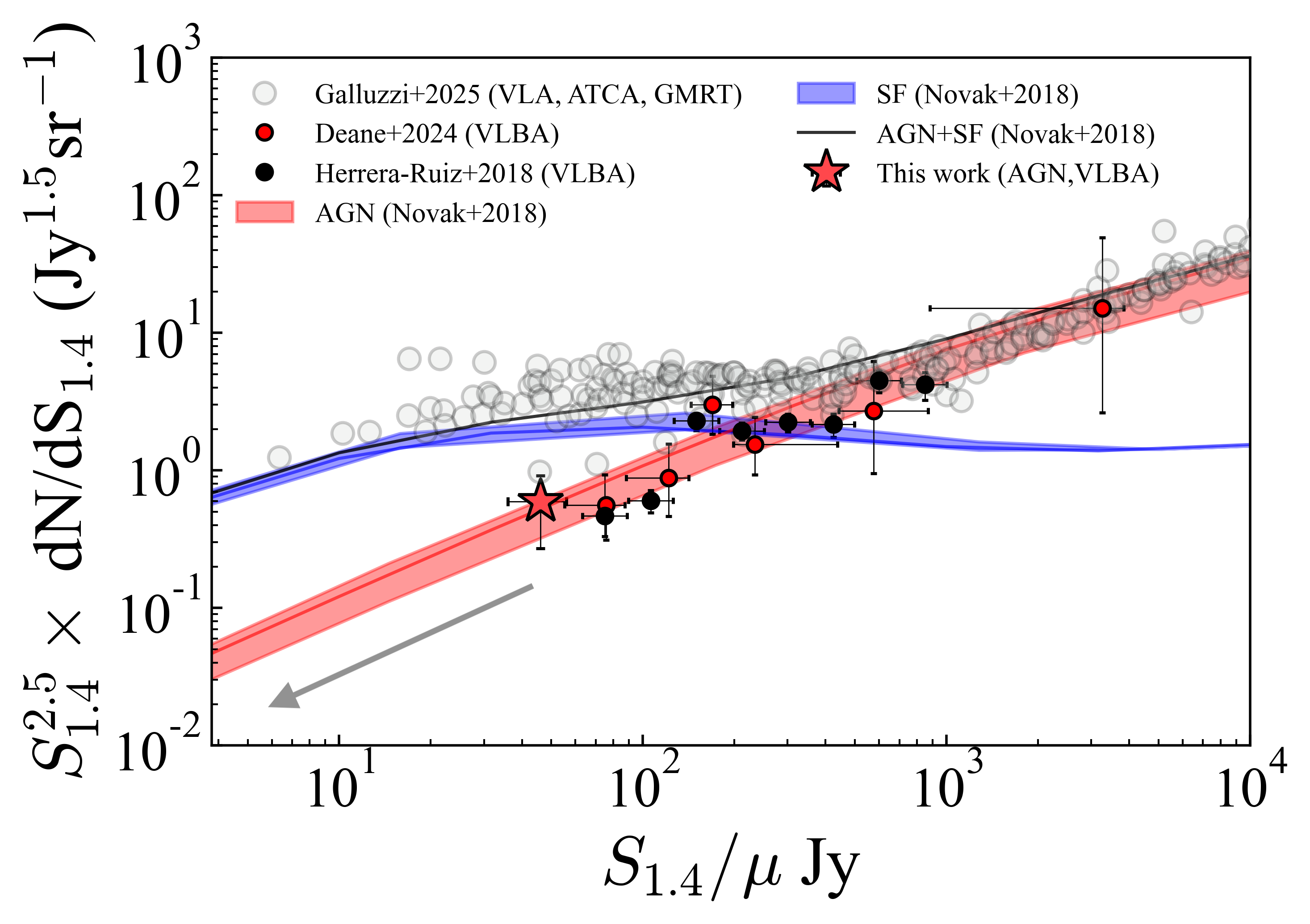}
    \caption{Euclidean normalized radio source number counts obtained by fitting, with the Markov chain Monte Carlo algorithm, the total radio Luminosity Function (LF) from \cite{novak+2018} using different evolving analytical LFs for AGN hosts (red line) and SF galaxies (blue line). The shaded areas encompass the 3$\sigma$ errors from the $\chi^2$ fits performed on individual populations \citep[see][for further details]{novak+2018}. Grey circles are the number counts from \cite{galluzzi+2025} obtained from a compilation of radio observations with different instruments (e.g. VLA 1-3 GHz, ATCA, GMRT, LOFAR, MeerKAT etc.), without any distinction between AGN or SF. Red and black circles show the number counts from the VLBA CANDELS GOODS-North Survey \citep{deane+2024} and VLBA 1.4 GHz COSMOS \citep{herrera-ruiz+2018} in case of radio emission from AGN activity. The red star is the number count of the detections from the AGN-sCAN survey presented in this work. The grey arrow point towards the regime below our sensitivity limit, from which the number counts extrapolated from the red curve of \cite{novak+2018} of the 42 brightest AGN were extracted, in order to constrain upper limits on our results (see text in Sec. \ref{sec:cumulative} for further details).}
    \label{fig:flux_density}
\end{figure*}

Following a similar approach as in \cite{deane+2024} and \cite{herrera-ruiz+2018}, we compute the differential source counts, dN/dS$_{\nu}$, as a function of source 1.4 GHz flux density, S$_{\nu}$, using the expression:

\begin{equation}
    \frac{dN}{dS} = \frac{N}{\Omega \Delta S}
\end{equation}

where $N$ is the number of sources in a flux density bin, $\Omega$ is the area over which a source with flux density equal to the mean flux density of the bin could be detected, and $\Delta$ S is the width of the flux density bin. \\
Figure \ref{fig:flux_density} displays the 1.4 GHz source count of the VLBA-detected AGN presented in this work (red star). For comparison, we also show the differential source counts derived by \cite{herrera-ruiz+2018} using the VLBA COSMOS survey (as black dots), the derived source counts for the VLBA CANDELS
GOODS-North Survey from \cite{deane+2024} (as red circles) and the VLA 3 GHz source counts split into SF and AGN subpopulations from \cite{novak+2018}. 
The \cite{novak+2018} curves are obtained by statistically separating SFGs and radio AGN based on the "radio-excess" criterion \citep{delvecchio+2017}, using data from the VLA-COSMOS 3 GHz Large Project \citep{smolcic+2017}.
Our results are perfectly consistent with these previous works, as indeed the VLBA source counts at $S_{1.4GHz}$ = 37-58 $\mu$Jy follow the red curve in \cite{novak+2018}, which is representative of the AGN population. At these fluxes, the AGN number counts are $\sim$5$\times$ lower than those of the total radio source population.  This corroborates that the average AGN fraction of $\sim$20\%  among sources within this flux regime, while $\sim$80\% of the sources are star-forming galaxies.

Our measurement in Fig. \ref{fig:flux_density} (red starred symbol) is the faintest source count measurement ever obtained from VLBI studies, and aligns well with previous (low-resolution) extrapolations of AGN number counts (red solid line). In this context, \cite{whittam+2017} hinted at a possible flattening of the AGN number counts towards faint flux densities, meaning that compact, flat-spectrum radio cores are increasingly dominant in these regimes. Although we do not have enough statistics to robustly test this hypothesis, we remark that the excess of flat-spectrum radio cores could indicate the presence of younger radio jets, short duty cycles, weak jet collimation, or slower jet speeds \citep[see][for an in-depth discussion]{deane+2024}.

\section{Discussion}

\subsection{Cumulative Distribution of AGN in the faintest regime } \label{sec:cumulative}

Our 5.5$\sigma$ sensitivity limit enabled us to detect AGN activity with flux densities down to $< S_{1.4} >$ = 37-58 $\mu$Jy. 

Of course, we cannot rule out a widespread contribution of AGN at even fainter fluxes. Unveiling those ultra-faint AGN is, however, very challenging. VLBA stacking would improve our sensitivity limit, but it is not feasible due to to the lack of prior knowledge on the AGN position. The VLA peak flux position was indeed not sufficient to serve as a prior, as indeed the VLBA has over $\times$10 times higher spatial accuracy. Similarly, HST or JWST positions are significantly less accurate than VLBA positions, with the additional caveat that they could trace star formation, which might not be spatially coincident with AGN emission. Therefore, we attempt a statistical assessment of the AGN contamination in VLBA non-detections.

We derive a quantitative threshold of how faint the required flux limit would have to be, by conservatively assuming that our VLBA non-detections from the (M$_\star$, rms)-matched CS are the first 42 brightest AGN expected just below the faintest flux of our VLBA detections, i.e., at S$_{1.4} <$ 37 $\mu$Jy.

To do so, we convert the Euclidean-normalized 1.4 GHz source counts into a dimensionless quantity, re-scaled to the effective sky area of the ($M_\star$,rms)-matched CS (i.e. 9.7 x 10$^{-6}$ sr). This yields a number of expected AGN in the same area, as a function of 1.4 GHz flux. 
We make sure that the relation between differential N$_{\rm AGN}$ and S$_{1.4}$ yields an integrated $N_{AGN}$ = 4 in the range 36--58 $\mu$Jy (i.e. matching our VLBA detections, red star in Fig. \ref{fig:flux_density}). Then we extrapolate the same trend down, and below our nominal VLBA flux density limit. Finally, we compute the cumulative distribution of $N_{AGN}$ at $S_{1.4} < 37 \mu$Jy, and we found out that 42 sources would be cumulated at a flux density of 5 $\mu$Jy, which is then the minimum flux density reached by the $\approx$ 42 brightest AGN. We stress again that this is an over-conservative assumption, since it is very unlikely that our non-detections are all clustered just below the flux limit. However, at these low flux densities, the brightness temperature drops to $T_b << 10^5$ K (see Eq. \ref{eq:t_b}). This suggests that, even if we were able to detect such faint sources with the VLBA, the origin of their emission would be ambiguous between AGN or SF.  Consequently, we conclude that pushing VLBA observation to sub-$\mu$Jy sensitivity would not be an effective strategy for revealing a residual hidden population of radio AGN.  For instance, according to the ngVLA exposure time calculator \footnote{\url{https://ngect.nrao.edu/}}, we note that Band 1 (2.3 GHz observations) would reach brightness temperatures T$_b \geq 10^5$ K (at 5$\sigma$, i.e. $\approx$ 5 $\mu$Jy/beam) in about 16h on-source. Hence, fainter fluxes, as expected for the bulk of our VLBA non-detections, would be detected with a lower T$_B<10^5$ K, thus carrying an ambiguous origin despite their easy detection.

Furthermore, even in this conservative scenario in which our non-detections are the brightest AGN below our sensitivity limit, the ratio $S_{VLBA}/S_{VLA}$ is 17\% at S$_{VLA}$ $\approx 70 \ \mu$Jy. $S_{VLBA}$ is the mean flux of the 42 brightest AGN obtained from the cumulative distribution of the AGN number counts in the range $ 5\ \mu{\rm Jy} \leq S_{1.4} \leq 36 \ \mu{\rm Jy} $, while $S_{VLA}$ is the mean VLA 1.4 GHz flux density of the 42 non-detections. This ratio would translate into a global upper q$_{IR}$ shift of $<$0.1 dex. We note that, based on the M$_{\star}$-dependent IRRC, $q_{IR}$ decreases by 0.22 dex within the same $M_\star$ range of the analysed sample \citep[$\log M_\star/M_\odot > 10.5$, ][]{delvecchio+2021}. Therefore, given that radio AGN contamination (including VLBA detections and non-detections) counteracts this effect by $< 0.1$ dex, the decreasing $q_{IR}$ with $M_\star$ is not primarily explained by AGN contamination.

Among other possible mechanisms that have been proposed to explain the decline of $q_{IR}$ at stellar masses $\log (M_\star/M_\odot) \geq 10.5$ in SFG, there is an increasing energy loss rate of high-energy CR electrons, which can be due to particularly efficient supernova-driven turbulence \citep{schober+2023}.
The CR energy loss could also depend on the gas surface density and the redshift of the host galaxy, as modeled by \cite{yoon+2024}. Finally,  \cite{ponnada+2025} found that in many cases "radio-excess" objects can be better understood as "IR-dimmed" objects with longer-lived radio contributions at low-$z$ from Type-Ia supernovae and intermittent black hole accretion in weakly SFGs.

\subsection{Diffuse AGN emission as a possible caveat}

A potential caveat of this analysis is that the AGN flux could be underestimated because of extended emission (beyond the VLBA beam, e.g. 200pc $\times$ 50pc at $z=1$), that is resolved out by the VLBA \citep[e.g.][]{morabito+2025a}. 
We first tested that the integrated fluxes remain consistent across various tapering and Briggs schemes in our images.
We also note that additional AGN flux over more extended physical scales would correspond to a lower T$_b$, hence $T_b <$10$^5$ K (see Eq. \ref{eq:t_b}). \\
A statistical assessment of this AGN emission at intermediate scales comes from \cite{herrera-ruiz+2017}.
The vast majority of their VLBA detections ($>$90\%) are classified as AGN based on the radio-excess observed from the VLA fluxes. Briefly, the "radio excess" is determined by the difference between the radio luminosity predicted by the IRRC at the redshift and stellar mass of the host galaxy (which traces star formation only) and the observed radio luminosity measured by VLA, which includes contributions from both star formation and AGN.
Assuming that this radio-excess traces the full AGN-related flux over the entire galaxy size, the comparison between total VLBA and AGN-driven VLA flux indicates how much flux is lost on larger scales. We note \citep[see Figure 11 in][]{herrera-ruiz+2017} that at fluxes S$_{1.4}$ $< 50 \mu {\rm Jy}$, the VLBA flux is already $>~ 75\%$ of the total VLA AGN flux, hence highly compact. By extrapolating this trend towards fainter flux regimes (as for our VLBA targets), we would expect that the VLBA flux is capturing the vast majority (if not all) of the total AGN flux.
Finally, the agreement between the predicted and observed number count in this work (Fig. \ref{fig:flux_density}) further supports the hypothesis that the VLBA effectively captures the full AGN emission from our targets.

\section{Conclusions}

We presented our 1.4~GHz VLBA observations carried out in the context of the AGN-sCAN survey (PI: Delvecchio), targeting a sample of 500 SFGs in COSMOS at z$>$0.5 located around the IRRC (see Fig. \ref{fig:qIRz0}). 
By exploiting this sample down to a limiting T$_B$$\gtrsim$10$^5$~K, we could test whether the decrease of $q_{IR}$ with galaxy stellar mass, as observed in \cite{delvecchio+2021}, could be explained by residual radio contamination from radio-faint AGN buried in the total radio emission of massive star-forming galaxies. To do so, we investigated the difference in the q$_{IR}$ distribution before and after correcting for a VLBA AGN contribution.

To perform a fair comparison between detections and non-detections, we built up a ($M_\star$,rms)-matched CS of 42 pure SFG, which did not show AGN emission as detected by the VLBA at 1.4 GHz and have stellar mass $\log (M*/M_\odot) \geq 10.5$.

Our findings can be summarized as follows:

\begin{itemize}
\item[(a)] We revealed a radio core in 4 targets placed in the deepest area. The effective AGN detection rate (rms-matched) in SFGs around the IRRC is $\sim$9\%. By subtracting the VLBA AGN contribution at 1.4 GHz, we find that the q$_{IR}$ distribution remains essentially unchanged. Indeed, the median and dispersion values of the distributions are entirely consistent with each other before and after applying this correction (as shown in the left and right panels of Fig. \ref{fig:deltaqir_mass}). Given that such AGN correction is negligible for $\log (M_\star/M_\odot) \geq 10.5$, we expect that it will be even less significant at lower stellar masses.

\item[(b)]  As shown in Fig. \ref{fig:flux_density}, the AGN number counts at 37--58$\mu$Jy lie well onto the extrapolation of the AGN number counts from \cite{novak+2018}, and the AGN contribution amounts to 20\% of the total number count of the entire radio population at these fluxes. Moreover, even in the extreme case that our VLBA non-detections were the brightest AGN just below the detection threshold, we still predict a maximum q$_{IR}$ shift of $<$ +0.1 dex. This upper limit is significantly smaller than the 0.22~dex decline of $q_{IR}$ within the M$_{\star}$ range of the VLBA detections, thus excluding the possibility that radio-AGN contamination is the primary driver of the M$_{\star}$-dependent $q_{IR}$.
\end{itemize}

Finally, we noted that pursuing deeper VLBA observations would not have been an effective strategy to unveil a residual hidden population of even fainter AGN. Other potential physical mechanisms that could explain the $q_{IR}$ decline in massive galaxies include CRe transport \citep{yoon+2024} and a longer-lived radio contribution from Type Ia supernovae \citep{ponnada+2025}. Nevertheless, our results highlight that the SFR-radio relation is robust against contamination from AGN activity, once the so-called "radio-excess" AGN are removed.

\begin{acknowledgements}

The authors thanks the referee for a constructive report that helped us clarify the results and implications of this work.
G.~P., I.~D., M.~G., F.~U., and I.~P. acknowledge funding by the European Union – NextGenerationEU, RRF M4C2 1.1, Project 2022JZJBHM: "AGN-sCAN: zooming-in on the AGN-galaxy connection since the cosmic noon" - CUP C53D23001120006. C.~S. acknowledges financial support from INAF $-$ Ricerca Fondamentale 2024 (Ob.Fu. 1.05.24.07.04), from the Italian Ministry of Foreign Affairs and International Cooperation (grant PGRZA23GR03), and by the Italian Ministry of University and Research (grant FIS-2023-01611, CUP C53C25000300001). 
IP also acknowledges support from INAF under the Large Grant 2022 funding scheme (project "MeerKAT and LOFAR Team up: a Unique Radio Window on Galaxy/AGN co-Evolution").
W.W. acknowledges the grant support through JWST programs. Support for programs JWST-GO-03045 and JWST-GO-03950 were provided by NASA through a grant from the Space Telescope Science Institute, which is operated by the Association of Universities for Research in Astronomy, Inc., under NASA contract NAS 5-03127.
This research has also been supported by the European Regional Development Fund under grant agreement PK.1.1.10.0007 (DATACROSS).
This work is based, in part, on observations made with the NASA/ESA/CSA James Webb Space Telescope. The data were obtained from the Mikulski Archive for Space Telescopes at the Space Telescope Science Institute, which is operated by the Association of Universities for Research in Astronomy, Inc., under NASA contract NAS 5-03127 for JWST. 
\end{acknowledgements}

\bibliography{biblio}{}
\bibliographystyle{aa}

\appendix

\section{Evolution of $q_{IR}$ with redshift} \label{sec:appendix-2}

\begin{figure}[h]
    \makebox[0.4\textwidth]{
    \centering
    \includegraphics[scale=0.85]{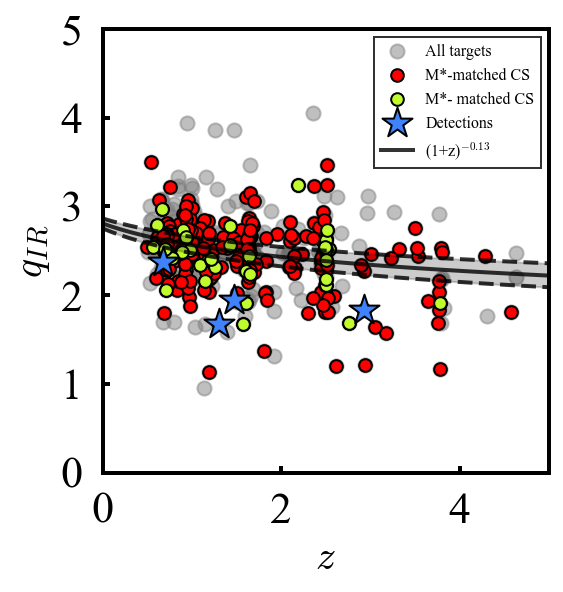}}%
    \caption{$q_{IR}$ as a function of redshift for all the 500 targets (grey points), the 4 VLBA-detected AGN (dark blue stars) and the 277 non-detections that constitute the $M_\star$-matched sample (red points). The green dots are the 42 non-detections which make the ($M_\star, rms$)-matched control sample. The black line is the NLS fit of all the targets described by the power law $(1 + z)^{-0.13}$.}
   \label{fig:qIRz0}
\end{figure}

In Figure \ref{fig:qIRz0} we show that $q_{IR}$ evolves with the redshift following a power law (1 + z)$^{A}$ (e.g. black line). This is because our targets were drawn from a flux-limited sample of joint IR and radio detections. From a non-linear squares (NLS) method, we find $A = -0.19 \pm 0.03$, in agreement with e.g. \cite{delhaize+2017}.
The same figure shows that our 4 VLBA detections (blue stars) are placed at different redshifts. The statistic is too low to explore the observed $q_{IR}$ distribution in multiple redshift bins, and for this reason, we decided to re-scale the $q_{IR}$ values to the same redshift, choosing a reference value $z_0$. 
We compute the $z$-corrected $q_{IR}$ using the expression:

\begin{equation}
    q_{IR, z_0} = q_{IR} (z) \ \left(\frac{1+z_0}{1+z}\right)^{A}
\end{equation}

where $z_0$ is the reference redshift, that we set to an arbitrary value of  $z_0 =0.5$. 
Finally, we show the 42 SFG (green dots in Fig. \ref{fig:qIRz0}) of the ($M_\star, rms$) - CS, by considering those VLBA observations in regions deep as or deeper than 6.7 $\mu {\rm Jy}$ beam$^{-1}$ (i.e., the poorest sensitivity reached among the detections). We find that the distributions of the mass-matched CS and the mass- and rms-matched CS spread randomly around the IRRC, indicating that systematic effects are absent.

\begin{figure}[h]
    \centering
    \makebox[-1\textwidth]{
    \includegraphics[scale=0.5]{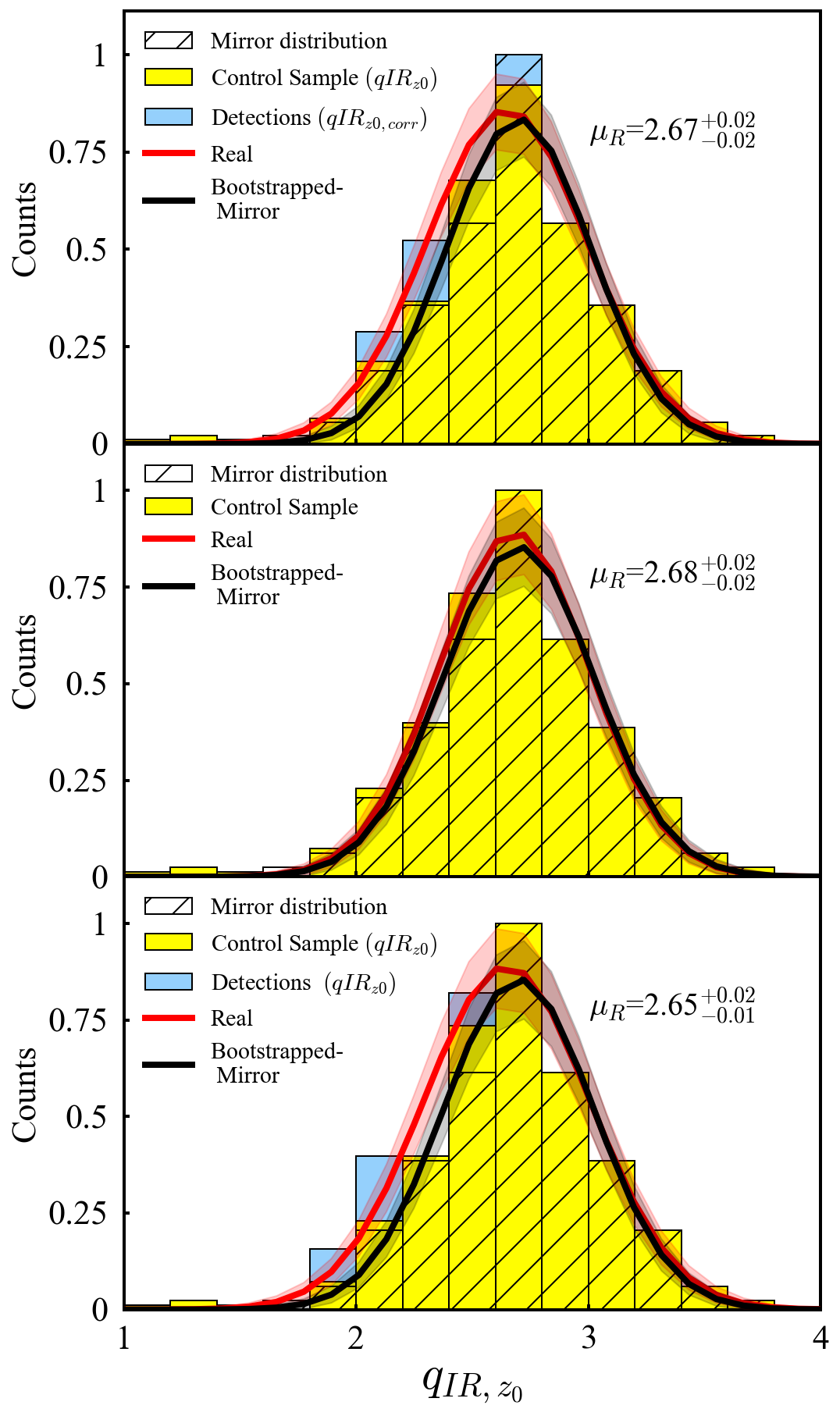}}%
    \caption{Normalized q$_{IR}$ distributions re-scaled to $z_0=0.5$ ($q_{IR,z_0}$) for the control sample of VLBA non-detections (yellow histogram), while the 4 detections (light blue histogram) are shown both before (bottom panel) and after the VLBA AGN correction (top panel). The red Gaussian is the fit of the yellow+blue (or yellow) histograms, while the black curve is the Gaussian fit to the bootstrapped mirror distribution.}
    \label{fig:distr_qIRz0}
\end{figure}

Then, we compute the corrected $q_{IR,z0}$ ($q_{IR,z_0,corr}$) of the 4 detections by subtracting the VLBA fluxes arising from the AGN, which is $\log (L_{1.4 GHz, VLA} - L_{1.4 GHz, VLBA})$. 
Figure \ref{fig:distr_qIRz0} shows the distribution of the $q_{IR, z_0,corr}$ of the 4 detections (blue histogram) and the $q_{IR,z_0}$ of the control sample (yellow histogram). The bin size is 0.2 dex, that is about twice larger than the average $q_{IR}$ uncertainty.

The red curve is the Gaussian fit of the real distribution.
The gray curve is obtained by extracting 1000 values from the so-called "mirror distribution" (hatched histogram) using a bootstrap method, which would represent the distribution observed if all galaxies were star-forming. The errors used in the bootstrap are the 16$^{th}$/84$^{th}$ percentiles of the 1000 values.
The mirror distribution is obtained by unfolding the right-hand side of the distribution (above the peak, which is assumed to originate from pure SF) on the left side.
A K-S test was not able to exclude that the mirror and the real distribution are drawn from different parent samples (with a p-value of 0.99). Accordingly, the median value of the real distribution, which is $\mu_{R}$ = 2.67 $^{0.02}_{-0.02}$, and the median of the bootstrapped-mirror distribution ($\mu_{\overline{M}}$ = 2.69 $^{0.04}_{-0.04}$), are consistent between the errors on the medians, computed as the 16$^{th}$/84$^{th}$ percentiles of the respective distributions divided by the root squared number of the total counts.
Thus, as expected, the real distribution matches the mirrored one after subtracting the AGN contribution from the $q_{IR,z_0}$.

To explore the effects of the correction from the AGN contribution, we study $q_{IR,z_0}$ in two different scenarios: 
(1) without the VLBA correction for both detections and non-detections; and (2) only for the control sample of non-detections. Interestingly, the real and mirrored distributions are still statistically indistinguishable in the (1) case. The bottom panel of Figure \ref{fig:distr_qIRz0}, indeed, shows that $\mu_{R}$ = 2.65 $^{0.02}_{-0.01}$ and $\mu_{\overline{M}}$ = 2.69 $^{0.04}_{-0.04}$, which are (again) consistent within the error bars between each other and with the values in Fig. \ref{fig:distr_qIRz0}. However, the difference between $\mu_{R}$ and $\mu_{M}$ is $\sim$ 0.04 dex and larger than when considering the AGN-corrected $q_{IR,z_0}$.
We also observe a slight shift on the left side between the red and black curves, reflecting the fact that all the detections have $q_{IR,z_0}$ $< \mu_{R}$. The shift, indeed, disappears in the central panel of Fig. \ref{fig:distr_qIRz0} when we consider SFG-only from the control sample. 
Finally, the values of $\mu_{R}$ are identical among the (1) and (2) cases.
In conclusion, we can robustly claim that VLBA AGN contamination, despite being non-negligible for individual detections, does not impact at all the peak position (and the overall shape) of the original $q_{IR}$ distribution.

\label{LastPage}

\end{document}